\documentclass[aps,prb,twocolumn,superscriptaddress,floatfix,showpacs]{revtex4-1}
\usepackage{tikz}
\usepackage{amsmath}
\usepackage{makeidx,amssymb,amsmath,amsthm,graphicx}
\usepackage[toc,page]{appendix}
\usepackage{xcolor}

\begin{document}


\title{Weak magnetic anisotropy in GdRh$_2$Si$_2$ studied by magnetic resonance.}

\author{J.Sichelschmidt}
\email[]{joerg.sichelschmidt@cpfs.mpg.de}
\affiliation{Max Planck Institute for Chemical Physics of Solids, 01187 Dresden, Germany}
\author{K.Kliemt}
\affiliation{Physikalisches Institut, Goethe-Universit\"at Frankfurt/M, 60438 Frankfurt/M, Germany}
\author{M.Hofmann-Kliemt}
\affiliation{Fachbereich Mathematik, Technische Universit\"at Darmstadt, 64289 Darmstadt, Germany}
\author{C.Krellner}
\affiliation{Physikalisches Institut, Goethe-Universit\"at Frankfurt/M, 60438 Frankfurt/M, Germany}
\date{\today}

\begin{abstract}
The antiferromagnetically (AFM) ordered state of GdRh$_{2}$Si$_{2}$ which consists of AFM-stacked ferromagnetic layers is investigated by magnetic resonance spectroscopy. The almost isotropic Gd$^{3+}$ paramagnetic resonance becomes anisotropic in the AFM ordered region below 107~K. The emerging internal anisotropic exchange-fields are still small enough to allow an investigation of their magnetization dynamics by using a standard microwave-frequency magnetic resonance technique. We could characterize this anisotropy in detail in the ferromagnetic layers of the excitation at 9 and 34 GHz. We derived a resonance condition for the AFM ordered phase to describe the weak in-plane anisotropic behaviour in combination with a mean-field analysis.
\end{abstract}

\maketitle

\def\neel{{N\'eel} }
\def\FA{F_{\rm an}}
\def\CA{C_{\rm an}}
\def\MS{M_{\rm sat}}
\def\EDF{E_{\rm df}  }
\def\text#1{{\rm #1}}
\def\i{\item}
\def\[{\begin{eqnarray*}}
\def\]{\end{eqnarray*}}
\def\bv{\begin{verbatim}}
\def\ev{\end{verbatim}}
\def\ganz{Z}
\def\3{\ss}
\def\reel{{\cal}R}
\def\platz{\;\;\;\;}
\def\beginvector{\left(\begin{array}{c}   }
\def\endvector{\end{array}\right)}
\def\fff{\frac{3}{k_B} F }
\def\vec#1{ {\rm \bf #1  } }
\def\KBMUEF{\frac{3k_B}{\mu_{\rm eff}^2 }}

\def\neel{{N\'eel} }

\def\FA{F_{\rm an}}
\def\CA{C_{\rm an}}
\def\MS{M_{\rm sat}}
\def\mueff{\mu_{\rm eff}}
\def\kbue{\frac{3k_B}{\mu_{\rm eff}^2}}
 
\def\unit{\hat{1}} 

\def\text#1{{\rm #1}}
\def\i{\item}
\def\[{\begin{eqnarray*}}
\def\]{\end{eqnarray*}}

\def\bv{\begin{verbatim}}
\def\ev{\end{verbatim}}

\def\ganz{Z}
 
\def\3{\ss}

\def\reel{{\cal}R}

\def\platz{\;\;\;\;}

\def\beginvector{\left(\begin{array}{c}   }
\def\endvector{\end{array}\right)}

\def\fff{\frac{3}{k_B} F }

\def\zudbs{\frac{3k_B}{\mu_{\rm eff}^2}}

\def\qT{\tilde{q}}
\def\pT{\tilde{p}}

\def\RT{\tilde{R}}
\def\QT{\tilde{Q}}
\def\PT{\tilde{P}}

\def\Bsf{B_{\rm sf}}

\def\aalpe{A_1}
\def\batem{A_2}


\section{Introduction}

%
GdRh$_{2}$Si$_{2}$ belongs to the silicides with tetragonal ThCr$_{2}$Si$_{2}$-structure which show exceptional magnetic properties, e.g. the antiferromagnetic Kondo systems YbRh$_{2}$Si$_{2}$ \cite{trovarelli00a} and CeRh$_{2}$Si$_{2}$ \cite{quezel84a}, HoRh$_{2}$Si$_{2}$ which exhibits so-called ``component separated'' magnetic transitions \cite{shigeoka11a} and a temperature tunable surface magnetism \cite{generalov17a}, or SmRh$_{2}$Si$_{2}$, showing unusual valence states of the Sm ions at the surface and in the bulk \cite{chikina17a}. GdRh$_{2}$Si$_{2}$ possesses antiferromagnetic (AFM) order of well localized magnetic moments appearing below $T_{\rm N}=107$~K \cite{kliemt15a} which is characterized by an AFM propagation vector (001) and a stacking of ferromagnetic layers \cite{kliemt15a,czjzek89a}. In spite of the pure spin ground state of Gd$^{3+}$ a weak in-plane anisotropy occurs which is indicated by the magnetization behavior of the ordered moments being aligned in the basal plane. A mean field model could describe the magnetization data with the assumption that the ordered magnetic moments are aligned parallel to the [110] direction \cite{kliemt17a}.\\
Recent angle-resolved photoelectron spectroscopy revealed two-dimensional electron states at the Si-terminated surface of GdRh$_{2}$Si$_{2}$ and their interplay with the Gd-magnetism. These surface states exhibit itinerant magnetism and their spin-splitting arises from the strong exchange interaction with the ordered Gd 4$f$ moments \cite{guttler16a}. \\
Magnetic resonance techniques are widely used to study the dynamic properties of magnetic ordering \cite{gurevich96a}. With GdRh$_2$Si$_2$ we study a prototypical material which not only exhibits a simple magnetic structure but also allows for the investigation of the magnetically ordered regime with conventional magnetic resonance techniques at low fields and frequencies.
We could estimate the anisotropy fields by applying a standard condition for the resonance modes in the ferromagnetic sublattices. However, for the resonance anisotropies observed in GdRh$_2$Si$_2$ common AFM resonance theories \cite{gurevich96a} turned out to be not applicable. Instead, we utilized a particular mean-field model for the AFM ordering to describe the angular dependence of the resonance field.

\section{Experimental}

High-quality single-crystalline GdRh$_{2}$Si$_{2}$ were used in this study. The growth and characterization of which is described in Ref. \onlinecite{kliemt15a}. We investigated the paramagnetic resonance (above $T_{N}$) and the magnetic resonance of the ordered moments (below $T_{N}$) by using a continuous-wave Electron Spin Resonance (ESR) spectrometer together with helium- and nitrogen-flow cryostats allowing for temperatures between 5 and 300~K. 
Two frequencies $\omega/2\pi=9.40$~GHz (X-band) and $\omega/2\pi=34.07$~GHz (Q-band) were utilized to evaluate the resonance field condition which in the paramagnetic region simply reads: $\omega/\gamma=H_{\rm res}$ where $\gamma=g\mu_B/\hbar$ is the gyromagnetic ratio and $g$ is the spectroscopic splitting factor.\\
In general, an ESR spectrometer allows to measure the absorbed power $P$ of a transversal magnetic microwave field as a function of a static and external magnetic field $\mu_0H$. A lock-in technique improves the signal-to-noise ratio by a field modulation which then yields the derivative of the resonance signal $dP/dH$ as the measured quantity. The resulting spectra were fitted with a Lorentzian function including the influence of the counter-rotating component of the linearly polarized microwave field \cite{rauch15a}. From the fit we obtained the resonance field $H_{\rm res}$ and the linewidth $\Delta H$ (half-width at half maximum). 

\section{Results and Discussion}
\subsection{Paramagnetic regime}

For the paramagnetic regime, i.e. for $T> T_{\rm N}=107$~K, the ESR spectra and their temperature dependence was discussed in a recent paper \cite{sichelschmidt17a}. The spectra display a behavior as typically expected for well-defined local moments in a metallic environment and with a temperature dependence as expected for anisotropic exchange-coupled paramagnets \cite{elschner97a,huber12b,kwapulinska88a}.  For temperatures nearby magnetic ordering the critical linewidth divergence could be described by a slowing down of in-plane ferromagnetic fluctuations within a model for a 3D Heisenberg ferromagnet \cite{benner90a}.
%
%
\subsection{Ordered regime: temperature dependence}
\label{ESRTdata}
\begin{figure}[hbt]
\begin{center}
\includegraphics[width=0.7\columnwidth]{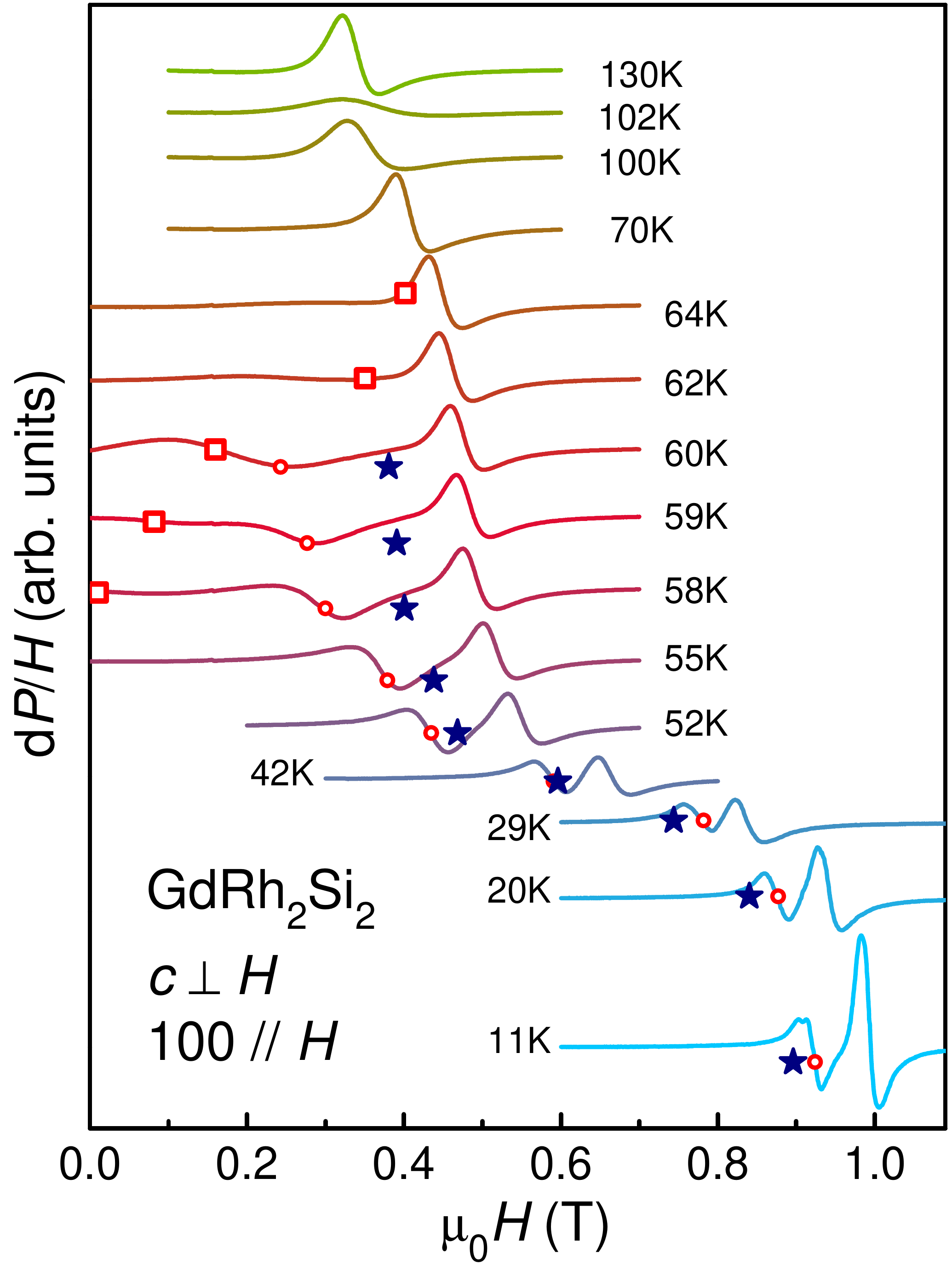}
\end{center}
\caption{
(Color online) X-band (9.4~GHz) magnetic resonance spectra at various temperatures, mostly in the magnetically ordered region ($T_{\mathrm N}=107$~K), for the external field along the particular in-plane direction (100). Open squares and circles indicate the resonance fields of additional lines at fields below the main line, see also corresponding symbols in \textcolor{blue}{Fig. \ref{FigOrdXQ}}. Stars indicate the spin-flop field as determined from magnetization data \cite{kliemt15a,kliemt17a}.
}
\label{FigSpecs100pH}
\end{figure}

GdRh$_{2}$Si$_{2}$ is a layered antiferromagnet below $T_{\rm N}=107$~K. The Gd 4$f$ moments are ferromagnetically ordered within the basal plane (with alignment parallel to the [110]-direction) while they stack in antiferromagnetic order along the [001]-direction \cite{kliemt17a}.\\
Figure \ref{FigSpecs100pH} shows selected spectra for the in-plane direction $H\|100$. Upon cooling below $T_{\rm N}=107$~K the paramagnetic resonance develops into a resonance mode of the magnetization of ferromagnetic (FM) sublattices. For temperatures below $\approx65$~K the spectra consist of more than two lines. The spectral structure indicated by open circles appears near the fields of the spin-flop transition (from magnetization data \cite{kliemt15a,kliemt17a}, indicated by stars). By sweeping across the spin-flop field the internal field rapidly changes and during this change it also matches the resonance condition (Eqn.~(\ref{dHaniso2})) which leads then to the observed structure. 
In a narrow temperature region between 55~K and 65~K a component of the easy-direction resonance ($H\|110$) is observed in the $H\|100$ -- spectra as indicated by the open squares. A slight misorientation might explain that. 

\begin{figure}[t]
\begin{center}
\includegraphics[width=0.7\columnwidth]{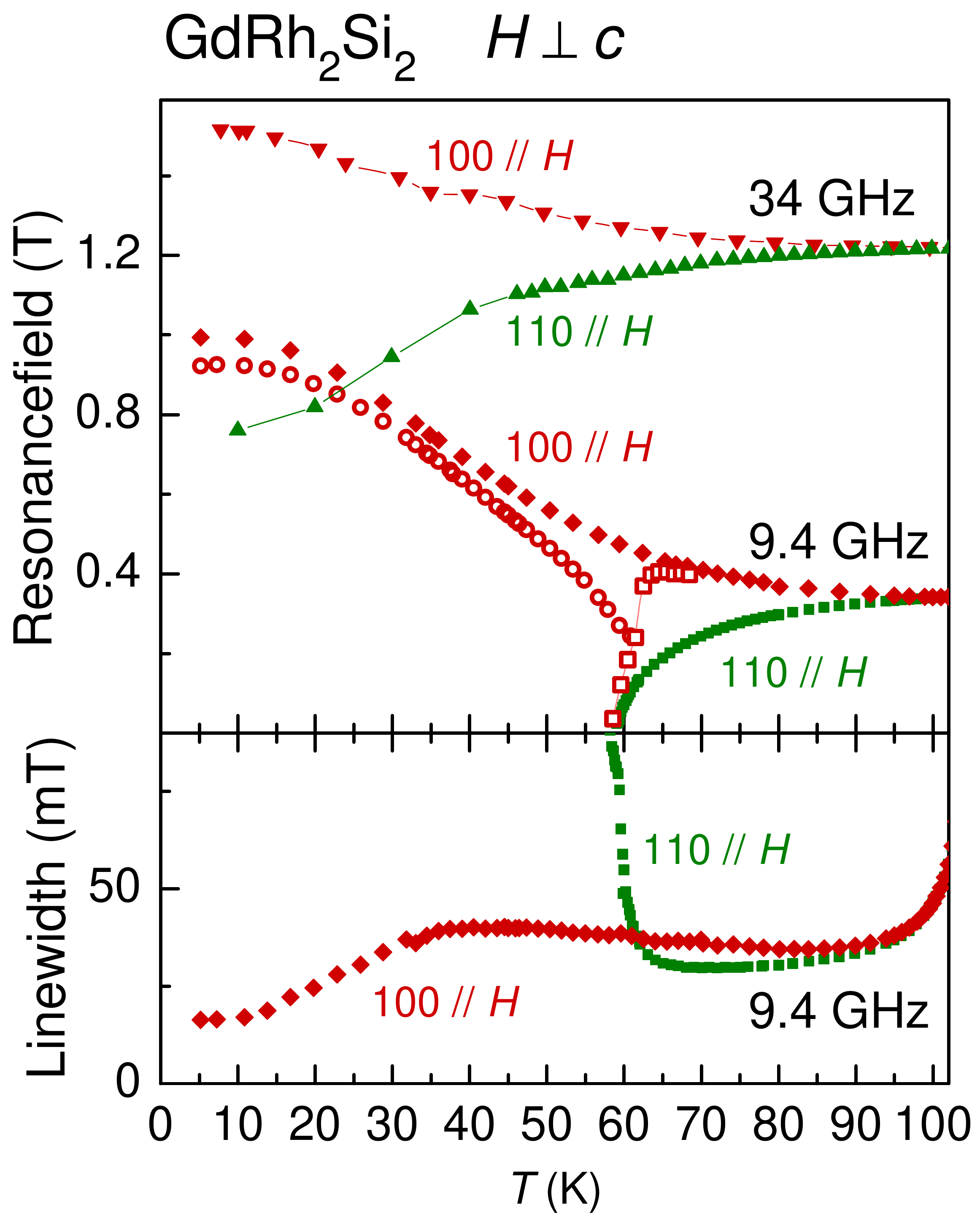}
\end{center}
\caption{
(Color online) Temperature dependence of resonance field $H_{\rm res}$ and linewidth $\Delta H$ for the external field along two different in-plane directions and two microwave frequencies as indicated. Solid lines guide the eyes. Open squares and circles indicate $H_{\rm res}$ of additional lines as shown in \textcolor{blue}{Fig. \ref{FigSpecs100pH}}. 
}
\label{FigOrdXQ}
\end{figure}

The spectral structures could be well described by Lorentzian lineshapes which results in resonance fields and linewidths as shown in Figure \ref{FigOrdXQ}. For the external field along the easy direction [110], the X-band spectra disappear at temperatures below about 60~K whereas the Q-band spectra are well defined down to the lowest temperatures. The reason for this behaviour is a temperature dependent anisotropy energy (field) which at $T=0$ is between the X- and Q-band energies (fields) and which matches the X-band energy at around 60~K.
Increasing the temperature towards $T_{\rm N}$ reduces the anisotropy of the line parameters, i.e. the anisotropy field decreases with increasing temperature. 

\begin{figure}[t]
\begin{center}
\includegraphics[width=0.7\columnwidth]{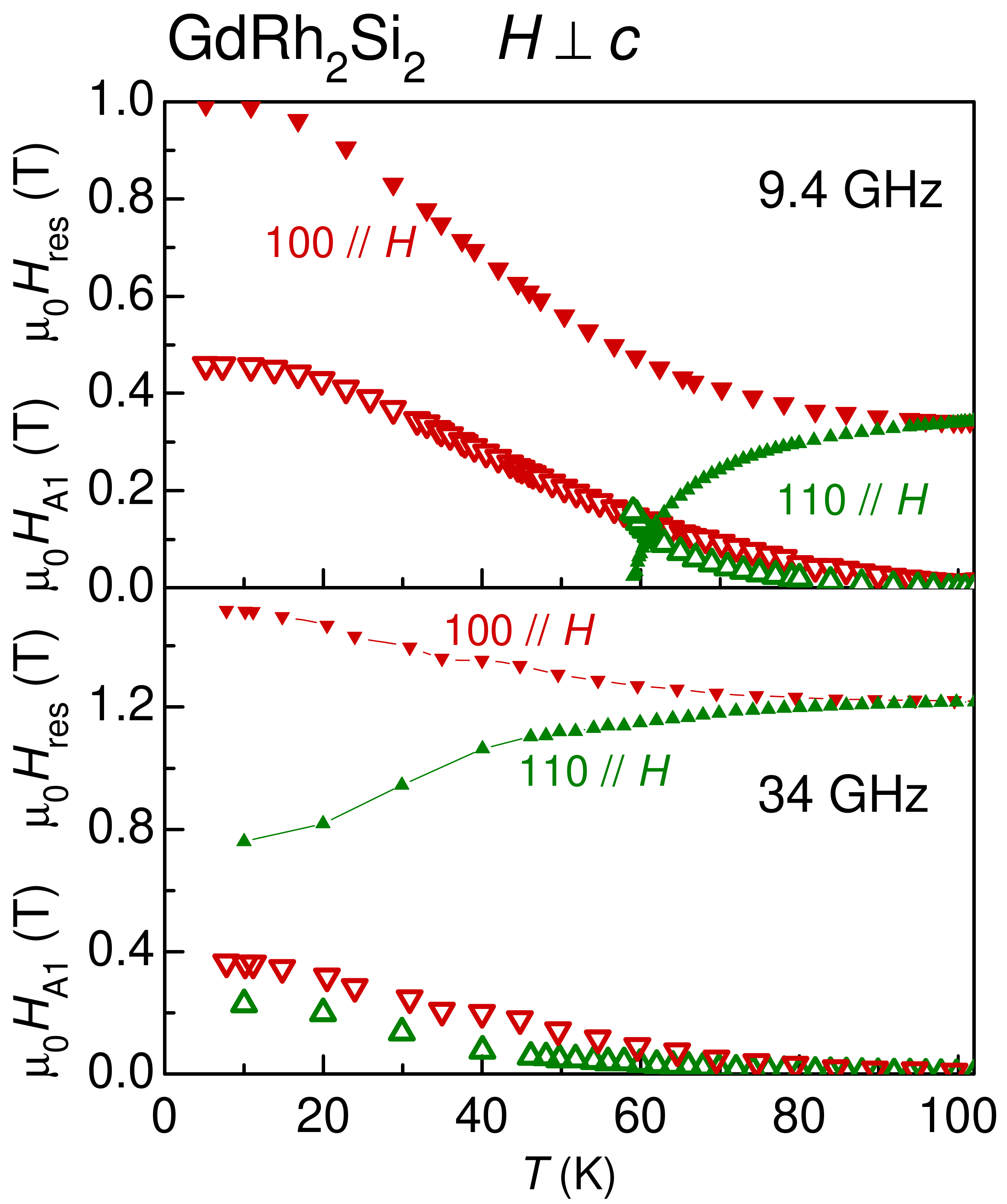}
\end{center}
\caption{
(Color online) Temperature dependence of resonance field $H_{\rm res}$ (closed triangles) and anisotropy field (open triangles, Eqns. \ref{HA1},\ref{HA2}) for the data at 9.4 and 34 GHz. 
}
\label{FigOrdHa}
\end{figure}
The anisotropy field can be estimated from the resonance field as follows. The conditions of a ferromagnetic resonance for a 
sample with cubic crystal structure may be used for an approach to describe the resonance fields in case of the ferromagnetic in-plane order in GdRh$_{2}$Si$_{2}$ \cite{gurevich96a} (for our case with the tetragonal in-plane anisotropy the symmetries are the same as those for the cubic case). With this, we get the resonance condition for a ferromagnetic sublattice:
\begin{flalign}
\mathrm{easy} & \,\mathrm{direction} \langle110\rangle: \nonumber\\
\omega/\gamma&=H_{\rm res}+2H_{\rm A1}\label{dHaniso1}\\[1ex]
\mathrm{hard} & \,\mathrm{direction}  \langle100\rangle: \nonumber\\ 
\omega/\gamma&=\left[\left(H_{\rm res}-2H_{\rm A1}\right)\left(H_{\rm res}+H_{\rm A1}+\frac{1}{2}H_{\rm A2}\right)\right]^{1/2}.\label{dHaniso2}
\end{flalign}
Here, $H_{\rm A1,A2}=K_{1,2}/M$ are anisotropy fields with $K_{1,2}$ first (second) order anisotropy constants. From Eqns. (\ref{dHaniso1},\ref{dHaniso2}) we calculated $H_{\rm A1}$, neglecting $H_{\rm A2}$: 
\begin{flalign}
\mathrm{easy} & \,\mathrm{direction} \langle110\rangle: \nonumber\\
H_{\rm A1}&=\textstyle\frac{1}{2}\left(\omega/\gamma - H_{\rm res} \right)\label{HA1}\\[1ex]  
\mathrm{hard} & \,\mathrm{direction}  \langle100\rangle: \nonumber\\ 
H_{\rm A1}&=-H_{\rm res}/4+\sqrt{\textstyle\frac{9}{16}H_{\rm res}^{2}-\textstyle\frac{1}{2}\left(\omega/\gamma\right)^{2}}. \label{HA2}
\end{flalign}
 Figure \ref{FigOrdHa} shows the results of Eqns. (\ref{HA1},\ref{HA2}) by using the experimental temperature dependent $H_{\rm res}$.

The anisotropy field $H_{\rm A1}$ has to be distinguished from the internal exchange fields which lead to magnetic order. The \textit{antiferromagnetic} order corresponds to an internal, in-plane exchange field which is much too large for an AFM resonance mode to be observable at GHz frequencies. 
According to Eqn.~(\ref{innerfieldFormel}) from App.~\ref{innerfield} the internal field which is created by the AFM stacked FM sublattices A and B is
\begin{equation}
B_{\rm interior,A,B}^{x}=\zudbs\Theta_{N}\sqrt{M_{sat}^{2}(1-\frac{T}{\Theta_{N}})
				-(\chi_{\perp}B_{z})^{2}}.
\end{equation}
With an external field component $B_{z}=0$ one obtains for $T\rightarrow0$, $M_{sat}=7\mu_{B}$, $\mu_{eff}=8.28\mu_{B}$, $\Theta_{N}=107$~K: 
$B_{\rm interior,A,B}^{x}=48.8$~T. Hence, in order to observe an antiferromagnetic resonance a resonance frequency of $\nu=g\mu_{B}/h\cdot B_{\rm interior,A,B}^{x}=1.37$~THz (g=2) is required. This may hard to be verified because THz spectroscopy requires samples with a good transmission for THz radiation - which is not the case for GdRh$_{2}$Si$_{2}$.

The in-plane \emph{ferromagnetic} order is caused by internal exchange fields allowing for the resonance observation at GHz-frequencies. 
The z-component of the internal field is solely determined by the external field $B_{z}$ as 
\begin{equation}
B_{\rm interior,A,B}^{z}=\zudbs\Theta_{W}\chi_{\perp}B_{z}
\end{equation}
again using  Eqn.~(\ref{innerfieldFormel}) from App.~\ref{innerfield}. One gets with $\chi_{\perp}(T=78\rm\,K)=0.1\mu_{B}/\rm T$ and $\Theta_W=8\,\rm K$
\begin{equation}
B_{\rm interior,A,B}^{z}/B_{z}=0.052.
\label{Bintz}
\end{equation}
This means that if an external field is applied along the $c$-axis only $\approx 5\%$ (at $T=78\rm\,K$) is internally available as an effective field for the magnetic resonance. For example,
using $B_{z}=6{\rm\,T}$ from an estimated value $\mu_{0}H_{\rm res}^{\|}=6{\rm\,T}$ of the out-of plane uniaxial resonance field (\textcolor{blue}{Fig.~\ref{AngleDep}}, left frame) one gets $B_{\rm interior,A,B}^{z}=0.31{\rm\,T}$. This value is close to the value for the X-band resonance field of Gd$^{3+}$ in the paramagnetic state \cite{sichelschmidt17a} and also close to the resonance field along the FM ordered direction [110]. 

%
\subsection{Ordered regime: anisotropy at 78~K}
\label{ESRAdata}
\begin{figure}[hbt]
\begin{center}
\includegraphics[width=1\columnwidth]{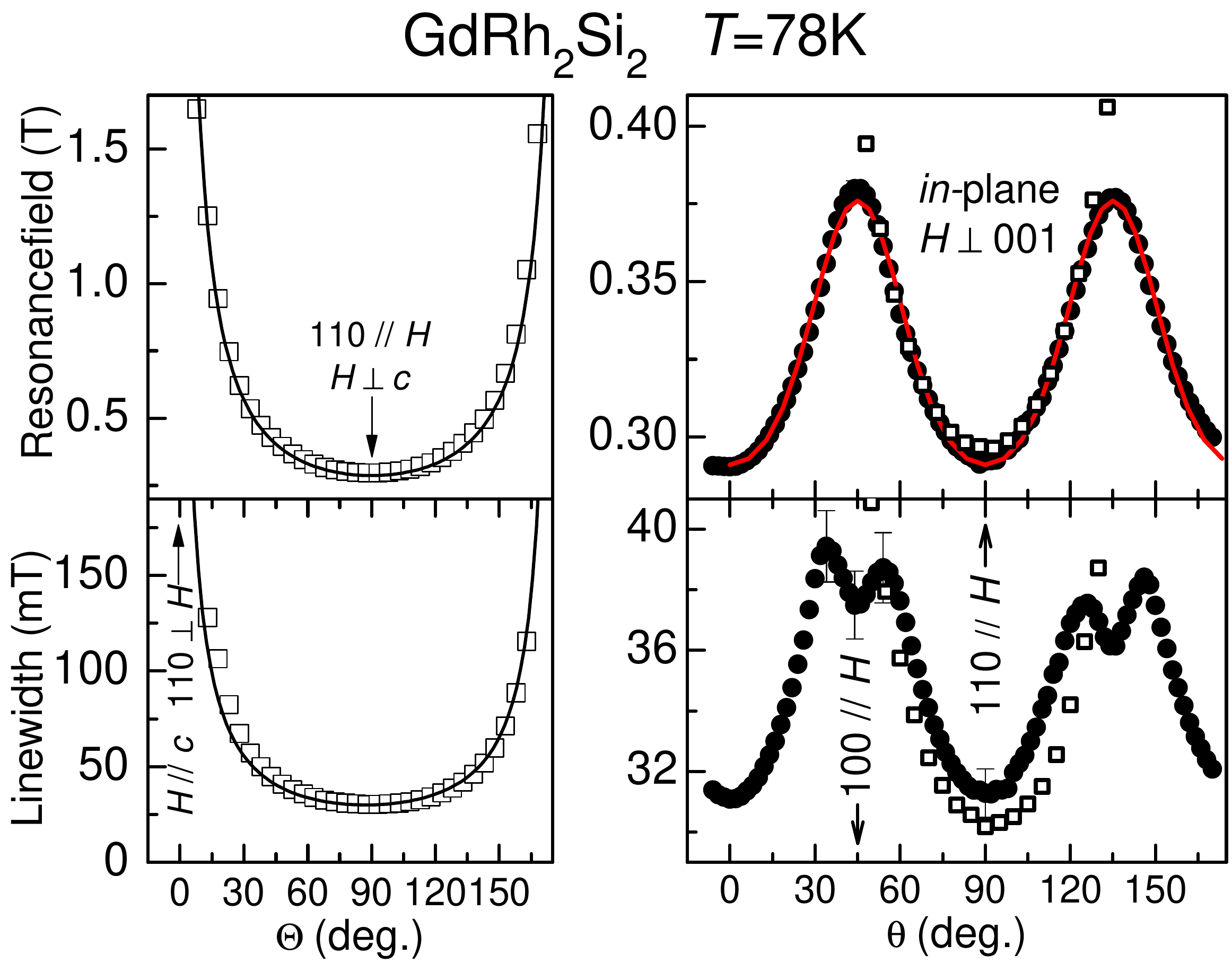}
\end{center}
\caption{
Angle dependence of X-band resonance field $H_{\rm res}$ and linewidth $\Delta H$, [110] is the easy direction of magnetization. External field is oriented by angles $\Theta$ and $\theta$ respective the indicated crystalline directions. {\it Left frame:} Out-of plane anisotropy. Solid lines indicate uniaxial behavior with $\mu_{0}H_{\rm res}^{\|}=6{\rm T}$, $\mu_{0}H_{\rm res}^{\perp}=0.29{\rm T}$ and $\mu_{0}\Delta H^{\|}=4{\rm T}$, $\mu_{0}\Delta H^{\perp}=0.03{\rm T}$.  {\it Right frame:}  In-plane anisotropy with external field $H$ in the basal plane (001) ($c\perp H$) at varying directions. Red solid line indicates Eqn.~(\ref{fitformel}) with $\xi=(\gamma_D /\gamma_M)^2\rightarrow0$ leading to $\gamma_D\rightarrow0$. Open squares indicate the data of the left frame.
}
\label{AngleDep}
\end{figure}

We investigated the anisotropy of the X-band data at $T=78\,\rm K$ where the linewidth for the 110 direction shows a minimum, see \textcolor{blue}{Fig.~\ref{FigOrdXQ}}. 
The anisotropy of resonance field and linewidth shown in \textcolor{blue}{Fig.~\ref{AngleDep}} is considerably stronger for tilting the external field out of the tetragonal plane (angle $\Theta$, left frame) than rotating it within the plane (angle $\vartheta$, right frame). Interestingly, the out-of-plane anisotropy can be nicely described by an uniaxial behavior (solid lines, left frame) just like a paramagnetic resonance with an uniaxial crystalline field anisotropy. This indicates that the effective internal field is always aligned along the external field. 
Also, the above internal exchange-field estimation, Eq.~\ref{Bintz}, shows that the value of the effective resonance field corresponds to a typical $g$-value of Gd$^{3+}$ as observed in the paramagnetic regime \cite{sichelschmidt17a}.


%

The in-plane anisotropy as shown in the right frame of \textcolor{blue}{Fig.~\ref{AngleDep}} presents a 90$^\circ$ periodicity of both resonance field and linewidth which reflects the fourfold symmetry in the tetragonal basal plane. The open symbols show the out-of plane data of the left frame. Obviously the angular dependencies of both in-plane and out-of-plane data sets are very similar near the easy direction of magnetization, $\langle110\rangle$. Such a behaviour can be understood as follows: 
Magnetization measurements at $T=78\,\rm K$ on single crystals yield a spin-flop field $B_{\rm sf}\approx 250 \rm{\,mT}$ for a field parallel to the $[100]$-direction and a domain-flip field $B_{\rm df}\approx 160 \rm\, mT$ for field parallel to the $[110]$-direction \cite{kliemt17a}. This implies that for fields of the order of the resonance field, applied along a main symmetry direction, the moments of both magnetic sublattices are in good approximation aligned perpendicular to that field (Figs.~3 and 5 in Ref.~\onlinecite{kliemt17a}). The magnetic moments can therefore be described as one large domain that extends over the whole single crystal.
%
%
Upon rotating the field in the basal plane away from a main symmetry direction, the magnetizations of the two sublattices are not equivalent anymore and a sine-like modulation of the resonance field occurs.
%

To model the in-plane behaviour and to describe the anisotropy in the ferromagnetic sublattices the solutions given by the standard theory for an AFM resonance \cite{gurevich96a} are not sufficient. We are not aware of any published approach which would be applicable to GdRh$_{2}$Si$_{2}$. Therefore, we derived an antiferromagnetic resonance condition for this anisotropy as described in Appendix B.  
The mean-field model that describes the magnetization of the system \cite{kliemt17a} together with the resonance condition Eqn. (\ref{omegasquared}) predicts the sine-like modulation with excellent quantitative consistency as is demonstrated by the red solid line in the right frame of \textcolor{blue}{Fig.~\ref{AngleDep}} which depicts Eqn.~(\ref{fitformel}) in App.~\ref{meanfield}.

The mean-field model \cite{kliemt17a} yields, that the values of $B_D$ and $B_M$ (see Appendix, Fig. \ref{ESR_Koord}) are different for different AFM domains. When approaching the [110]-direction, the energy difference between both domains decreases and according to the domain distribution estimated by the Ising chain model \cite{kliemt17a} both domains coexist. On the other hand, by approaching the [100]-direction, the predicted values of $B_D$ and $B_M$ become almost equal in value such that the magnetic resonance frequency of both domains becomes similar, too.  
The structure in the angle dependence of the linewidth around $\vartheta=45^{\circ}, 135^{\circ}$ may be therefore due to a superposition and exchange-narrowing of anisotropic resonance signals arising from different domains. A similar behaviour was suggested for CdCr$_{2}$S$_{4}$ where four resonance fields are combined via exchange narrowing into one line \cite{ehlers12a}.

\section{Summary}

GdRh$_2$Si$_2$ presents an exemplary case for easy-plane magnetic order with a weak in-plane magnetic anisotropy. The presented magnetic resonance data in the magnetically ordered regime depict a ferromagnetic resonance mode displaying features similar to a paramagnetic resonance mode. Its anisotropy can nicely be described by a resonance condition for the ferromagnetic sublattices with weak anisotropy together with a mean-field model which assumes that the ordered magnetic moments are aligned parallel to the [110]-direction \cite{kliemt17a}.
%
\begin{acknowledgments}
KK and CK gratefully acknowledge support by the DFG through grant KR3831/5-1. We acknowledge helpful discussions with Christoph Geibel, Hans-Albrecht Krug von Nidda and Zhe Wang. We are particularly grateful to Dieter Ehlers for his generous help and interest.
\end{acknowledgments}

\appendix
\setcounter{equation}{0}

\section{Internal field in the AFM phase}\label{innerfield}

In Ref. \onlinecite{kliemt17a} a free energy-based model to describe the AFM phase 
of GdRh$_2$Si$_2$ was introduced. 
The free energy is
\[
F&=&-TS - \frac12\,(\vec{M}_A+\vec{M}_B)\cdot\vec{B} 
         + \phi(\vec{M}_A,\vec{M}_B)
\\
&=&-TS - \frac12\,(\vec{M}_A+\vec{M}_B)\cdot\vec{B}
+E_{\rm FM} +E_{\rm AFM} +F_{\rm an} 
\]
with the contribution within a plane
\[
E_{\rm FM}
=-\zudbs\,(\Theta_W + \Theta_N\,)\,\frac18\,(\,\vec{M}_A^2+\vec{M}_B^ 2\,)
\]
and between planes
\[
E_{\rm AFM}
=-\zudbs\,(\Theta_W - \Theta_N\,)\,\frac14\,(\,\vec{M}_A\cdot\vec{M}_B\,)
.\]
The anisotropic part $F_{\rm an}$ will be neglected
for the discussion of the c-direction. 
We consider the field that is produced by a ferromagnetic plane B
and acts on an ion of the sublattice A 
\[
F&=&
\underbrace{-\frac12 \vec{M}_A \cdot \vec{B}}_{\rm Zeeman\, Term}
- 
\underbrace{\zudbs \,(\Theta_W - \Theta_N\,)
\,\frac14\,(\,\vec{M}_A\cdot\vec{M}_B\,)
}_{\rm between\, layers}
+ \cdots \\
&=&
-\frac12\, \vec{M}_A \big[\vec{B}
+
\underbrace{\zudbs \,(\Theta_W - \Theta_N\,)\,\frac12\,\vec{M}_B\,)
}_{\vec{B}_{\rm interior,B}}\big]+ \cdots.
\]
In the following, we determine the magnetization $\vec{M}_B$ of one ferromagnetic layer.
We have
\[M^2+D^2=D_0^2,\quad\quad
D_0=M_{\rm sat}\sqrt{1-\frac{T}{\Theta_N}},\]
with $M_{\rm sat}=7\,\mu_{\rm B}.$
For the field along the c-direction we have
$\vec{M}\perp \vec{D}$ and in particular
\[
M(B)=\chi_\perp B_z.
\]
This results in 
\[
\vec{M}_A=(D,0,M)=(\sqrt{D_0^2 - (\chi_\perp B_z)^2}   , 0  ,\chi_\perp B_z   )
\]
\[\text{and}\quad
\vec{M}_B=(-D,0,M).
\]
Therefore we have
\[
\vec{M}&=&\frac12(\vec{M}_A+\vec{M}_B)
=(0,0,M),\\
\vec{D}&=&\frac12(\vec{M}_A-\vec{M}_B)
=(D,0,0).\\
\]
For the choice of the coordinate system see Fig.~8 in Ref.~\onlinecite{kliemt17a}.
The field that acts on an ion of the sublattice A, which 
is created by the sublattices A and B, reads
\begin{eqnarray}
&&\vec{B}_{\rm interior,A,B}(T,B_z)
\\
&=&
-2\,\frac{\partial}{\partial\vec{M}_A}\,[E_{\rm FM} + E_{\rm AFM}]
\nonumber\\
&=&
\zudbs
\Big[\Theta_W\,\frac12\,(\vec{M}_A+\vec{M}_B)
+\Theta_N\,\frac12\,(\vec{M}_A-\vec{M}_B) \Big]
\nonumber\\
&=&
\zudbs
\left\{
\frac{\Theta_W}{2}\,\vec{M}
+
\frac{\Theta_N}{2}\,\vec{D}
\right\}
\nonumber\\
&=&
\zudbs
\begin{pmatrix} 
\Theta_N\,\sqrt{M_{\rm sat}^2 (1-T/\Theta_N)-(\chi_\perp B_z)^2} \\ 
0 \\ 
\Theta_W\,\chi_\perp B_z
\end{pmatrix}.
\label{innerfieldFormel}
\end{eqnarray}
The values of  $\Theta_W=8\,{\rm K}$ and  $\Theta_N=107\,{\rm K}$
and $\mueff=8.28\,\mu_{\rm B}$ and $\chi_\perp=0.149 \mu_{\rm B}/{\rm T}$
have been determined by magnetic measurements \cite{kliemt15a,kliemt17a}.


\section{Inplane anisotropy}
\label{meanfield}
\begin{figure}
\centering
\includegraphics[width=0.3\textwidth]{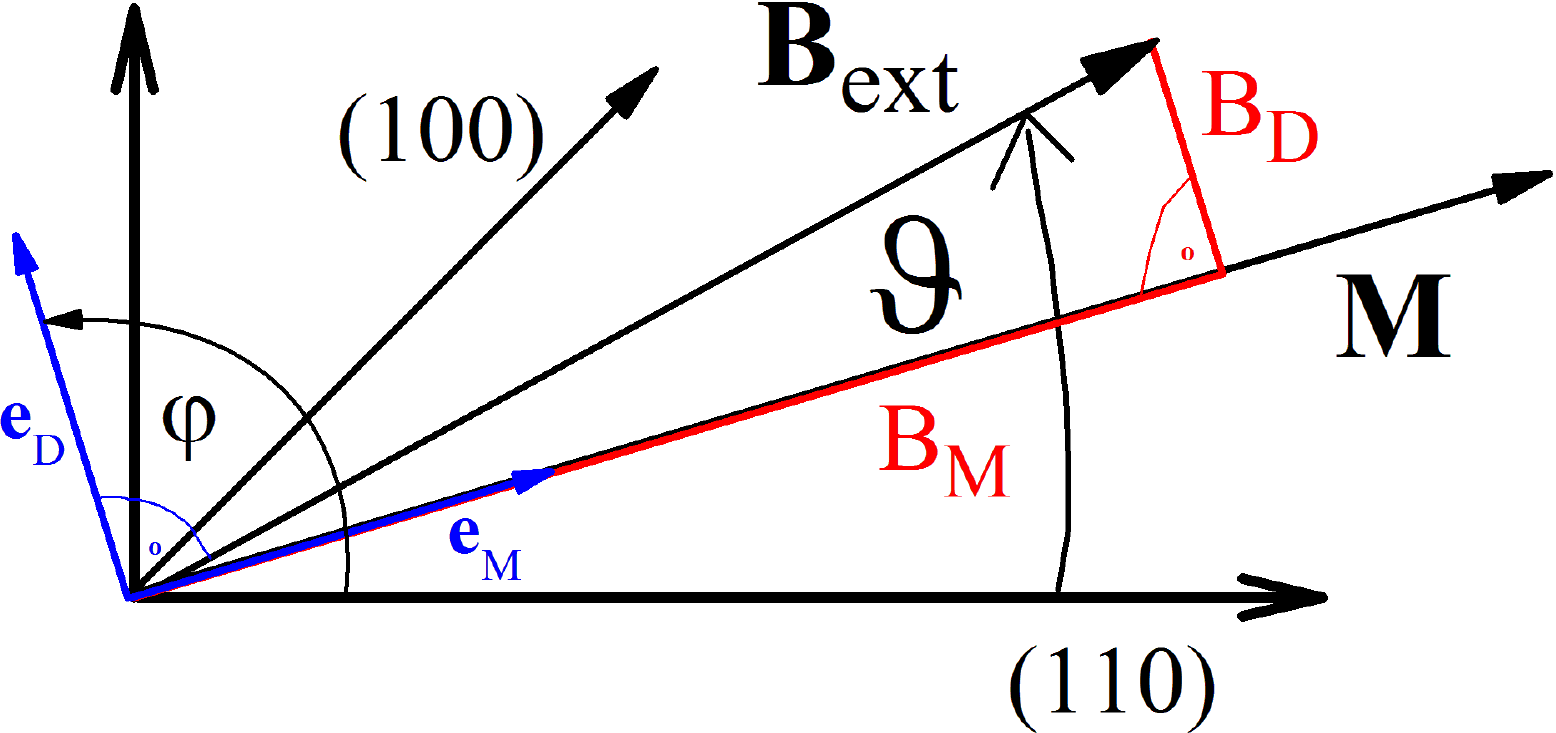}
\caption[]{On the choice of the coordinate system.}
\label{ESR_Koord}
\end{figure}

In the following, we consider the behaviour of one domain. We use the meanfield model developed 
in Ref.~\onlinecite{kliemt17a} to predict the ESR resonance field for an external field $\vec{B}$ applied perpendicular to the crystallographic $[001]$-direction.
The free energy 
\[
F(\varphi)\,
=\,F_0(D_0)&&
-\frac{B^2}{4}(\chi_\perp + \chi_\parallel)
-\frac{B^2}{4}(\chi_\perp -\chi_\parallel)\,u
\\
&&
+\frac{B_{\rm sf}^2}{8}\,(\chi_\perp -\chi_\parallel)\,\sin^22\varphi
\]
with
$u=-\cos(2\vartheta - 2\varphi)$,
Eqn.~(6) \cite{kliemt17a}, and the magnetization 
\[
\vec{M}=
\hat{\chi}\vec{B}
=\chi_\perp\,(1-\vec{e}_D\otimes\vec{e}_D)\,\vec{B}
+\chi_\perp\,\vec{e}_D\otimes\vec{e}_D\,\vec{B},
\]
Eqn.~(4) \cite{kliemt17a} serve as the starting point. 
For the choice of the coordinate system see Fig.~\ref{ESR_Koord}.
For our purpose it is sufficient to ignore $\chi_\parallel$.
The ESR interaction is that fast, that we do not expect 
an isothermic relaxation.
Here, only $\chi_\perp$ is relevant since this keeps the entropy unchanged.
Therefore, the magnetization becomes
\[
\vec{M}=
\chi_\perp\,B_M\vec{e}_M
\]
with $B_M=\vec{B}\cdot\vec{e}_M$ for $\vec{B}$ aligned 
perpendicular to the $[001]$-direction.
The free energy 
\[
&&F(\varphi)
 \\
&=&F_0 -\chi_\perp \,\frac{B^2}{4}\, [\,1-\cos(2\vartheta - 2\varphi)\,]
+\chi_\perp \,\frac{B_{\rm sf}^2}{8}\,\sin^22\varphi
\]
can be minimized with respect to $\varphi$
\[
&&\frac{\partial}{\partial\,\varphi}\, F(\varphi)
\\
&=&\chi_\perp \,\frac{B^2}{2}\,\sin(2\vartheta - 2\varphi)
+\chi_\perp \,\frac{B_{\rm sf}^2}{4}\,\cos 2\varphi\,\sin 2\varphi
\\
&=&0.\]
With
$B_D=B\,\cos(\vartheta - \varphi)$
and $B_M=-B\,\sin(\vartheta - \varphi)$
the minimum condition reads
\begin{equation}
-B_D\,B_M+\frac14\,B^2_{\rm sf}\,\sin\,4\varphi=0.
\label{meanminimum}
\end{equation}
The decomposition of $B=B_{ext}$ (Fig. \ref{ESR_Koord}) into $B_{D}$ and $B_{M}$ is done with respect to one AFM domain which consists of two FM sublattices.
In the paramagnetic regime, 
the ESR frequency $\omega$ can be decomposed in the following way:
The square of  $\omega$ is the sum
of three parts that arise from 
the three components of the external field.
This reads as
\[
\omega^2=\omega_x^2+\omega_y^2+\omega_z^2
\]
with 
\[
\omega_x=\gamma\,B_x, \platz 
\omega_y=\gamma\,B_y, \platz
\omega_z=\gamma\,B_z.
\]
We deduce a similar ansatz,
to describe the ESR frequency in the ordered regime.
In particular, we describe the ESR behaviour of one certain domain.
First of all, we introduce $\omega_{\rm field}$ and take into account
that the external magnetic field
 decomposes in a parallel $B_D$
and orthogonal component $B_M$ with respect to the ordering parameter $\vec{D}$. 
Furthermore, we account for the anisotropy in the system by utilizing $\omega_{\rm aniso}$ 
and add a constant term $\omega_0$.
These considerations lead to the ansatz
\[
\omega^2=\omega_0^2
+\omega_{\rm aniso}^2+\omega_{\rm field}^2(\vec{B})
.\]
An arbitrary analytic function
that respects the symmetry of one domain has the form
\[
\phi(\vec{B})=\phi_0 + c_D\,B_D^2 + c_M\,B_M^2
\]
and we can write
\[\omega_{\rm field}^2(\vec{B})
= \gamma^2_D\,B_D^2 + \gamma^2_M\,B_M^2.\]
For symmetry reasons, there 
is a $\pi/2$ periodicity upon rotations in the basal plane of the tetragonal lattice, such that
\[
\omega_{\rm aniso}^2=\omega_{\rm an}^2\,\cos\,4\varphi
\]
and summation yields
\begin{eqnarray}
\omega^2=\omega_0^2
+\omega_{\rm an}^2\,\cos\,4\varphi
+ \gamma^2_D\,B_D^2 + \gamma^2_M\,B_M^2
\label{omegasquared}
\end{eqnarray}
for the resonance frequency.
To introduce the amplitudes into the fit formula,
we use the $[100]$-direction where the resonance field has its maximum
$B_{\rm max}$ and the $[110]$-direction 
where the resonance field has its minimum $B_{\rm min}$.
We choose the coordinate system such that $\varphi$ is
the angle between the $[110]$-direction and the ordering vector $\vec{D}$.
In both cases, the external field is parallel to a main symmetry axis,
such that $B=B_M$.
This leads to 
\[
\omega^2&=&\omega_0^2
-\omega_{\rm an}^2+ \gamma^2_D\,B_D^2 + \gamma^2_M\,B_{\rm max}^2
\\
\omega^2&=&\omega_0^2
+\omega_{\rm an}^2+ \gamma^2_D\,B_D^2 + \gamma^2_M\,B_{\rm min}^2.
\]
From these two equations we determine
\[
\omega^2-\omega_0^2
&=& \gamma_M^2\,\frac12\,(\,B_{\rm max}^2+B_{\rm min}^2\,)
\\
\omega_{\rm an}^2
&=& \gamma_M^2\,\frac12\,(\,B_{\rm max}^2-B_{\rm min}^2\,)
.\]
With Eqn.~(\ref{omegasquared}) we get
\begin{eqnarray}
&&\frac12\,(\,B_{\rm max}^2+B_{\rm min}^2\,)
-\frac12\,(\,B_{\rm max}^2-B_{\rm min}^2\,)\,\cos\,4\varphi
\nonumber
\\
&=&\xi\,B_D^2+ B_M^2
\label{amp}
\end{eqnarray}
with a parameter $\xi=(\gamma_D /\gamma_M)^2$ to be determined by fitting.
To parametrize the plot in 
$\varphi$, which is
the angle between the x-axis ($[110]$-direction) and the
direction of the ordering vector $\vec{D}$,
we rewrite Eqn.~(\ref{meanminimum}) and Eqn.~(\ref{amp}) and get
\begin{eqnarray}
 B_M^2
+\xi
\,B_D^2&=&\aalpe \label{lau} 
\\
B_M\,B_D&=&\batem\label{magau}
\end{eqnarray}
with
\begin{eqnarray}
\aalpe&:=&\frac12\,(\,B_{\rm max}^2+B_{\rm min}^2\,)
\nonumber \\
&&-\frac12\,(\,B_{\rm max}^2-B_{\rm min}^2\,)\,\cos\,4\varphi
\label{A1}
\end{eqnarray}
and
\begin{eqnarray}
\batem:=\frac14\,B^2_{\rm sf}\,\sin\,4\varphi.
\label{A2}
\end{eqnarray}
To solve these equations we multiply Eqn.~(\ref{lau}) by $B_M^2$ and get a quadratic equation
\[
B_M^4+\xi\,\batem^2
=\aalpe\,B_M^2.
\]
From the two solutions we use the larger one
\[
B_M^2=\frac{\aalpe}{2} + \sqrt{\frac{\aalpe^2}{4} -\xi\,\batem^2}
\]
such that
$| B_D | < | B_M |$ is fulfilled.
Now we compute the component of the external field that is parallel to $\vec{D}$
\[
B_D^2=\frac{\batem^2}{B_M^2}
=
\frac{1}{\xi}
\left\{
\frac{\aalpe}{2} - \sqrt{\frac{\aalpe^2}{4} -\xi\,\batem^2}
\right\}.
\]
This yields the square of the external field for the resonance condition:
\begin{eqnarray}
&&B_{\rm res}^2=
B_M^2+B_D^2 \nonumber
\\
&=&
\left(\frac12 +\frac{1}{2\xi} \right)
\aalpe
+
\left(\frac12 -\frac{1}{2\xi} \right)
\sqrt{\aalpe^2 - 4\,\xi\,\batem^2}.
\label{Bsquared}
\end{eqnarray}
Since we have
\[
B_M=B_{\rm res}\,\cos(\vartheta-\varphi),\platz 
B_D=B_{\rm res}\,\sin(\vartheta-\varphi)
\]
we get
\[
\batem=B_M\,B_D=\frac12\,B^2_{\rm res}\,\sin(2\vartheta-2\varphi).
\]

With this we get a relation between the resonance field $B_{\rm res}$ and 
the angle $\vartheta$ between the external field and the $[110]$-direction 
\begin{eqnarray}
\vartheta=
\varphi+\frac12\,{\rm arcsin}\,\frac{2\batem}{B_{\rm res}^2}
\label{fitformel}
\end{eqnarray}
using Eqns.~(\ref{A1}),(\ref{A2}),(\ref{Bsquared}).
For $\xi\rightarrow0$ (and $\gamma_D\rightarrow0$) the ESR data are described well by Eqn.~(\ref{fitformel}) as shown in Fig.~\ref{AngleDep}, right frame.

\bibliography{JoergBib}

\end{document}